\documentclass[12pt,preprint]{aastex}

\newcommand{\kmag}{K$_{\mathrm S}$}
\newcommand{\jk}{J$-$K$_{\mathrm S}$}
\newcommand{\absk}{M$_{\mathrm K}$}
\newcommand{\irc}{IR~C~star}

\shorttitle{Age of Milky Way Bar}
\shortauthors{Cole \& Weinberg}

\begin{document}

\title{An Upper Limit to The Age of the Galactic Bar
}

\author{Andrew A. Cole \& Martin D. Weinberg}
\affil{Department of Astronomy, 
University of Massachusetts, Amherst, MA, USA 01003-9305;
{\it cole@condor.astro.umass.edu, weinberg@astro.umass.edu\\
\vspace{0.2in}
Accepted for publication in The Astrophysical Journal Letters}}

\begin{abstract}
Using data from the Two Micron All-Sky Survey (2MASS), we
identify a population of infrared carbon (IR~C) stars with \jk\ 
$\geq$ 2 in the Milky Way.  These stars are shown to trace
the stellar bar previously identified in IR and optical surveys.
The properties of C~stars strongly suggest that they are of 
intermediate age.  We conclude that the bar is likely to have
formed more recently than 3~Gyr ago, and must be younger than
6~Gyr.  Implications and further tests of this conclusion are
briefly discussed.
\end{abstract}

\keywords{Galaxy: structure --- stars: carbon --- Galaxy: disk}

\section{Introduction}

It is by now well-established that there is a strong 
stellar bar in the inner disk of the Milky Way.
A central bar was first hypothesized by \citet{dev64},
although the observational evidence has only become overwhelming
during the past decade (e.g., Kuijken 1996). 
Near and far-infrared source count maps have led numerous
authors to identify a triaxial bar roughly 3--5 kpc long,
with its near side in the first quadrant of the Galaxy
\citep[e.g.,][]{bli91,nak91,wei92,wei94}.  The OGLE survey
extended these maps of the inner Galaxy to optical wavelengths
and also found the bar morphology \citep{sta94,sta97}.  Recent
work based on 2MASS data \citep{skr01} shows the full bar quite clearly.

Dynamical arguments suggest that bars in galaxies may be
triggered by satellites and companions, intrinsic halo
asymmetry, or as a result of disk instability.  Once formed,
bars may be important for driving the evolution of galaxies
through global angular momentum redistribution and increased
rates of gas transport to galaxy centers.  As a result, barredness
may affect a galaxy's star-formation history and nuclear activity.
The dynamics of bars together with knowledge of bar ages may be
crucial to understanding disk-halo interactions and merger histories.
Despite the wealth of data on bar morphologies, little
is known of their ages, except for 
the statistical result that bar frequency
appears to decline with redshift for $z \gtrsim 0.5$ \citep{abr99}.
If the stellar populations of the Milky Way bar can be age-dated,
then the approximate time of the triggering event can 
be established.  This is a first
step towards reconstructing the dynamical history of disk
asymmetries in the Galaxy.

We know that infrared carbon (IR~C) stars reliably map out the
Milky Way bar \citep[][this paper]{skr01}.  
In \S2, we briefly recapitulate the
selection criteria for \irc s that allow us to identify this
population and to invert its magnitude distribution into
a distance distribution.  \S3 describes the
arguments for an intermediate age for the \irc s.  We argue
that they are certainly younger than 6~Gyr, and probably
younger than 3~Gyr, and therefore the bar must have formed
more recently than these limits.  In \S4, we discuss the
implications of this result for studies of bar formation
and suggest further work to refine our conclusion.  In short,
if the Milky Way is typical, bars might be temporary features
that can be successively re-excited during a galaxy's lifetime.

\section{Tracing the Bar with 2MASS}

Carbon stars come in several flavors, differing in their colors
and luminosities, evolutionary states, kinematics, and surface
enrichment patterns.  The brightest ($-$4 $>$ M$_{\mathrm{bol}}$
$>$ $-$6), and reddest (\jk\ $\gtrsim$ 1.2) C~stars are the classical
N-type AGB stars; they are typified by IRC~$+$10216 (CW~Leo).
Throughout this paper, the term ``C~star'' will be used
to refer to these cool, luminous AGB stars only; the term ``IR~C~star'' will
be used to refer to the subset of C~stars redder than (\jk)$_0$ = 2.

The \irc s may be identified, with care, in the 2MASS
(\jk, \kmag) CMD \citep{skr01}.
To illustrate, Figure~1a shows the Hess diagram of a 
4 deg$^2$ region of the Large Magellanic Cloud (LMC).  A
detailed exploration of the 2MASS view of the 
LMC has been
made by \citet{nik00}, so we will only touch on the most
prominent features here.  The main ridgeline of stars
extending up to \kmag\ $\approx$ 11 is the red giant and
asymptotic giant (AGB) sequence of the intermediate-age and old
stellar population.  The brighter plume reaching \kmag\
$\approx$ 8 consists of red supergiants.  The C~stars
occupy a bilinear sequence redward of \jk\ $\approx$ 1.4, 
with a slope change at \jk\ $\approx$ 2.  The redder 
sources are intrinsically similar to the warmer stars,
but are enveloped in a thick shroud of dust.  Thus the
extended C~star sequence lies nearly parallel to the 
reddening vector.   

The \irc\ branch of the 2MASS CMD of the Milky Way is shown
in Figure~1b.  The figure includes roughly 3 $\times$10$^4$
sources drawn from the entire sky excluding the Magellanic 
Clouds.  The color-magnitude distribution is wider than that
in Figure~1a because of the large distance spread within the
Galaxy and the high number of deeply embedded sources in
the larger Galactic sample.
An additional dispersion due to contamination by OH/IR stars
must be present; these have similar luminosities and near-IR
colors to the \irc s \citep{lep95}.

Spectra of stars in this region of the CMD have been obtained
in the LMC \citep[e.g.,][]{hug91}, the outer Galaxy
\citep[e.g.,][]{lie00}, and the inner Galaxy (M. Skrutskie, 
private communication).   A high fraction of the measured stars
have C-type spectra, with significant contamination by
highly reddened (E(B$-$V) $> 2$) M~supergiants and OH/IR stars.
We thus expect that a high fraction of the total number of
color-selected sources are carbon-rich too.  

Hipparcos parallaxes have shown $\langle$\absk $\rangle$ =
$-$7.6 $\pm$ 1 for carbon-rich Mira and SRa variables
\citep{wal98}.
The LMC population has
$\langle$\kmag $\rangle$ = 10.8 and $\sigma _K$ = $\pm$0.6, which 
yields $\langle$\absk $\rangle$ = $-$7.7 for an assumed distance
modulus of 18.5 mag.  We are therefore confident that the Galactic
sample is similar to the LMC \irc s.  Figure~1 shows
that the \kmag\ magnitude of the \irc s, with an appropriate color
correction, is stable enough to allow rough calculations
of Galactic structure.  A least-squares fit to the 
\irc\ branch of the LMC yields a linear relation, 
$\Delta$\kmag\ = $-$0.48~(\jk\ $-$ 2).  Application of this shift
to the individual \kmag\ magnitudes reduces all \irc s to a common
basis from which their approximate line of sight distances can
be derived.  The result is shown in Figure~2, which shows the
symmetric disk plus central bar morphology of the Galaxy.

Our simple inversion distorts the image, although 
the bar signature is unambiguous.  The width of the \absk \ 
distribution, due to pulsational variations and age/metallicity
differences, stretches and twists the bar in Figure~2.  
We systematically account for these biases by using a 
Bayesian parameter estimation for an exponential disk with
a quadrupole bar.  Restricting the sample to $|b|>2\arcdeg$\
to minimize confusion error, and correcting for extinction
\citep{sch98}, we find the most likely disk parameters to be:
a scale length $a=3.5\pm0.15$ kpc, a bar radius of $2.6\pm0.15$ kpc,
a position angle $\phi=31\pm8\arcdeg$, and a relative bar strength
of $0.32\pm0.15$.  The method and results will be further described
in a later paper.  The position angle and shape of the
inferred bar are hatched in Figure~2.  Density cuts through the
data along lines parallel and perpendicular to the Sun-Galactic
center line are shown for comparison to the exponential scale
length in Figure~3.

The low sensitivity of IR colors to reddening ensures that the 
\irc\ sequence is relatively free from contamination for colors
redder than \jk\ $\approx$ 2.5.  Even if large differential 
reddening is present, its effect should be small because the 
\irc\ sequence runs virtually parallel to the reddening vector.
The striking symmetry of our inferred Galactic disk shows that
selection effects are not driving our result.

\section{The Ages of Carbon Stars}

Whether a thermally pulsing AGB star becomes a C~star depends on 
its initial mass and metallicity.  Theoretical calibration
of the minimum mass (and hence maximum age)
necessary for C~star formation is uncertain due to the poorly understood
physics of the dredge-up process \citep{mar99}.  Empirical determination
of the age range for C~star formation has been hampered by the lack of 
massive ($\gg10^3$ M$_{\sun}$) 
intermediate-age star clusters in the galaxy,
and the age gap in Large Magellanic Cloud clusters between 3 and 9 Gyr
\citep{mar96}.  Synthetic models put the lower mass limit
for C~star production at 1.13 M$_{\sun}$ for Z = 0.004, and 1.32 M$_{\sun}$
for Z = 0.008; these masses correspond to ages of 5.2 and 3.9 Gyr, 
respectively \citep{mar99}.
Empirical evidence confirms the theoretical prediction for a
diminishing probability of C~star formation with increasing metallicity
\citep{coo86}.

\citet{cla87} determined the scale height of field C~stars to be 
similar to that of main-sequence F~stars,
which have ages of roughly 1--3~Gyr.
Star cluster ages are more easily measured;
Figure~4 shows a sample of clusters
in the Milky Way and the Magellanic Clouds whose C~star
content is known.  The clusters are sorted by age and 
M$_{\mathrm V}$.  
Solid and open symbols differentiate between clusters with
and without \irc s.  For clusters with C~stars, the approximate
number is given in parentheses, with the total C~star population
before the slash and the \irc\ population, where present, after.  
C~star content is taken from \citet{fro90}, \citet{nis00}, and
\citet{sca79} for the open and Magellanic
clusters.  Globular clusters do not contain C~stars 
\citep{mcc85,wal98}.  While many clusters
aged younger than 2~Gyr containing
\irc s have been omitted from Figure~4, we are unaware of {\it any} clusters
older than 3.2~Gyr containing-- or suspected of containing-- them.

The oldest cluster in the Galaxy thought to contain 
an \irc\ is Trumpler~5
\citep{kal74}.  Trumpler~5 is aged 2.8 Gyr, and membership of
the \irc, V493~Mon, is not confirmed due to the lack of proper 
motion studies of the cluster.  NGC~2121, 
an LMC cluster aged 3.2~Gyr, is the oldest cluster known
to harbor an \irc.  Metal-poor SMC clusters as old as 5--7 Gyr
contain luminous C~stars, but none as red as \jk\ = 2.  The efficiency
of C~star production dramatically drops with age among bright
clusters: compare
the C~star content of NGC~419 (1.6~Gyr, 9 C~stars)
with NGC~416 (6.9~Gyr, 1 C~star), and the globular cluster NGC~362
(12~Gyr, no C~stars).

Perhaps \irc s are born, but with reduced efficiency, at 
ages larger than 3~Gyr; the lack of bright clusters aged 3--5~Gyr
makes it difficult to tell.  A stricter limit
may be imposed by the lack of \irc s in M32.   M32 is known to harbor
a significant intermediate-age stellar population \citep{oco80};
\citet{del01} derive 3--5 Gyr and Z $\approx$ Z$_{\sun}$.  However,
\citet{cor01} find that its JHK colors are dominated by early K~giants,
and \citet{dav00} reports that M32's AGB terminates
blueward of J$-$K = 1.5.  Because of the high luminosity of M32 
(M$_{\mathrm V}$ = $-$16.7) compared to
a star cluster, its blue AGB may be the strongest observational
evidence for a cutoff in C~star production at approximately 3 Gyr for
populations of roughly Solar metallicity.

In summary, all empirical age estimates for \irc s indicate ages
less than about 3 Gyr.  This agrees with theoretical expectations
for a shutdown of the C~star formation mechanism between
4--5 Gyr for stars of Magellanic Cloud metallicity, and possibly younger 
ages for higher metallicities.  The oldest known C~stars are roughly
6~Gyr old,  but the metallicity dependence of C~star formation
and the lack of very red AGB stars in M32 makes ages of 1--3~Gyr
much more probable.

\section{Discussion and Conclusions}

We have used 2MASS data to trace the structure of the Galactic
disk and bar using the \citet{skr01} sample of \irc s
with (\jk ) $>$ 2.
The epoch of strong star formation
along galactic bars is expected to be brief 
\citep[$\lesssim$1~Gyr---][]{mar95,mar97}, and to occur during
their formation.  Little or no star formation is expected to 
occur within the bar once it has become well-established, 
and stars that subsequently form in the disk cannot themselves
join the bar.  Some support for this view comes from the 
observation that H {\small II} regions are common in the 
bars only of late-type, active, or morphologically disturbed
galaxies \citep[e.g.,][and references therein]{mar97}.
Therefore, the progenitors of the \irc s were born
prior to or during the formation of the bar, and hence
their lifetimes give an upper limit to the bar age.

The oldest attested 
C~stars (in SMC clusters with [Fe/H] $\approx -1.3$)
are 5--7~Gyr old.   If all the C~stars in the Galactic bar
were similar to the metal-poor SMC stars, reddened into
our color window, the bar could thus be as old as 
roughly 6~Gyr.  Among the oldest star clusters with \irc s
are NGC~2121 (LMC, 3.2~Gyr), and {\it possibly} Tr~5 (Milky
Way, 2.8~Gyr).  M32 appears not to contain \irc s \citep{dav00},
and contains a large stellar population aged 3--5 Gyr
\citep{del01}.  We thus find it highly probable that the Milky 
Way bar is younger than 3~Gyr.  This time frame is intriguing,
given the reports of a declining bar frequency among galaxies
with redshifts $z \gtrsim 0.5$ \citep{abr99}, corresponding
to lookback times of 5--6~Gyr.

Few age estimates have been made for the Milky Way bar.  \citet{ng96}
ascribed an 8--9~Gyr stellar population in Baade's Window
to the bar, but the population's spatial distribution is
not known, making the bar identification tentative.  \citet{sev99}
inferred an age of 7.5~Gyr from OH/IR stars with 
M $\approx$ 1.3~M$_{\sun}$; updated theoretical models
\citep{gir00} give 4.7~Gyr for this mass\footnote{Because 
the OH/IR stars yield a similar age to the \irc s, even
a catastrophic misjudgment of the C~star fraction in Figure~1b
does not invalidate our bar age limit.}.  The main-sequence turnoff
of a 3~Gyr-old population should be readily traceable along
the Galactic bar from V $\approx$17 at the near end to 
V $\approx$19 at the far end (modulo reddening differences).

Could the \irc s be a trace population that wandered into the
bar by chance?  
The LMC has
M$_{\mathrm{disk}}$ $\approx$ 10$^{10}$ M$_{\sun}$, and contains
2100 \irc s.  \citet{sta97} give M$_{\mathrm{bar}}$ $\approx$ 
2 $\times$ 10$^{10}$ M$_{\sun}$, and we find 5300 sources with
\jk\ $>$ 2 inside the bar radius.  For reasonable rates of
contamination by OH/IR stars, the bar seems to have an \irc\
specific frequency comparable to the LMC's, arguing against
the idea that 2MASS is seeing a trace young population within
a much older bar.  2MASS observations of the LMC show that our
color cut excludes roughly 75\% of the N-type C~stars 
\citep{nik00}.  If more of the C~stars could be traced, an
accurate estimate of the fraction of intermediate-age stars
in the bar could be made \citep{aar85}.

Could the \irc s be older than roughly 3~Gyr?  
Membership studies for Tr~5 and other potential 
C~star-bearing open clusters would help to empirically
set the upper age limit for C~star formation.  
Further study of M32's AGB, to definitively measure
its red extent, could push the likely age of the bar
upwards, if its \jk\ reaches $\approx 2.5$ \citep{fre92}.

Bar dust lanes can have high molecular gas content,
yet low star-formation rates \citep[e.g.,][]{dow96}.
Even in galaxies undergoing transient bursts (e.g., NGC~7479), 
the intensity of star formation along the bars is generally
suppressed relative to that of the disk \citep{lai99}.
This suggests that a bar contains a 
snapshot of the disk stellar population at the time of its formation.
However, barred galaxies typically have enhanced star formation
at the ends of their bars and in their nuclei.  Could these young 
stars join the bar as they form?  A star can only be trapped in 
the bar by losing angular momentum as its orbit passes through 
a resonance.  This can only happen efficiently if the bar is 
actively evolving.  The existence of a well-described molecular
ring in the Galaxy, presumably near corotation, 
suggests that our bar has a stable pattern speed, and therefore 
is not rapidly swallowing its own ends.

Bars are ubiquitous
in numerical simulations; detailed studies by many
authors show that the quasistatic gas response agrees well
with observed morphology \citep[e.g.,][]{ath92}.  This 
morphological coincidence suggests that the bars are 
at least several rotation times ($\sim$1~Gyr) old, and 
that their pattern speeds are not rapidly evolving.
On the other hand, a concentrated dark matter halo can
cause a bar to lose angular momentum through dynamical
friction.  \citet{deb00} have argued that bars such as 
the Milky Way's are strongly braked by this effect, 
which would imply rapid bar evolution.
However, \citet{wei02} have hypothesized that a primordial
bar that forms during the epoch of disk assembly will torque
up the inner halo as it is braked, producing a core in 
the halo density distribution.  This initial, transient
bar paves the way for a long-lived, stellar bar, since
the altered halo mass profile does not slow bars as efficiently.
The genesis of the current Milky Way bar remains to be explained.
Tidal triggering by the LMC, the Sagittarius dwarf, 
or a now-merged satellite is a plausible origin
\citep{mur98,wei98,ves00}.

\acknowledgments

We thank Neal Katz and Mike Skrutskie for comments on the
manuscript, and Imants Platais for advice about open clusters.
This publication makes use of data products from the Two Micron
All Sky Survey, which is a joint project of the University of 
Massachusetts and IPAC/Caltech, funded by NASA and the NSF.
This work was supported in part by NSF award AST-9988146.
This research has made use of the SIMBAD database, operated
at CDS, Strasbourg, France.

\clearpage

\begin{figure}
\plotone{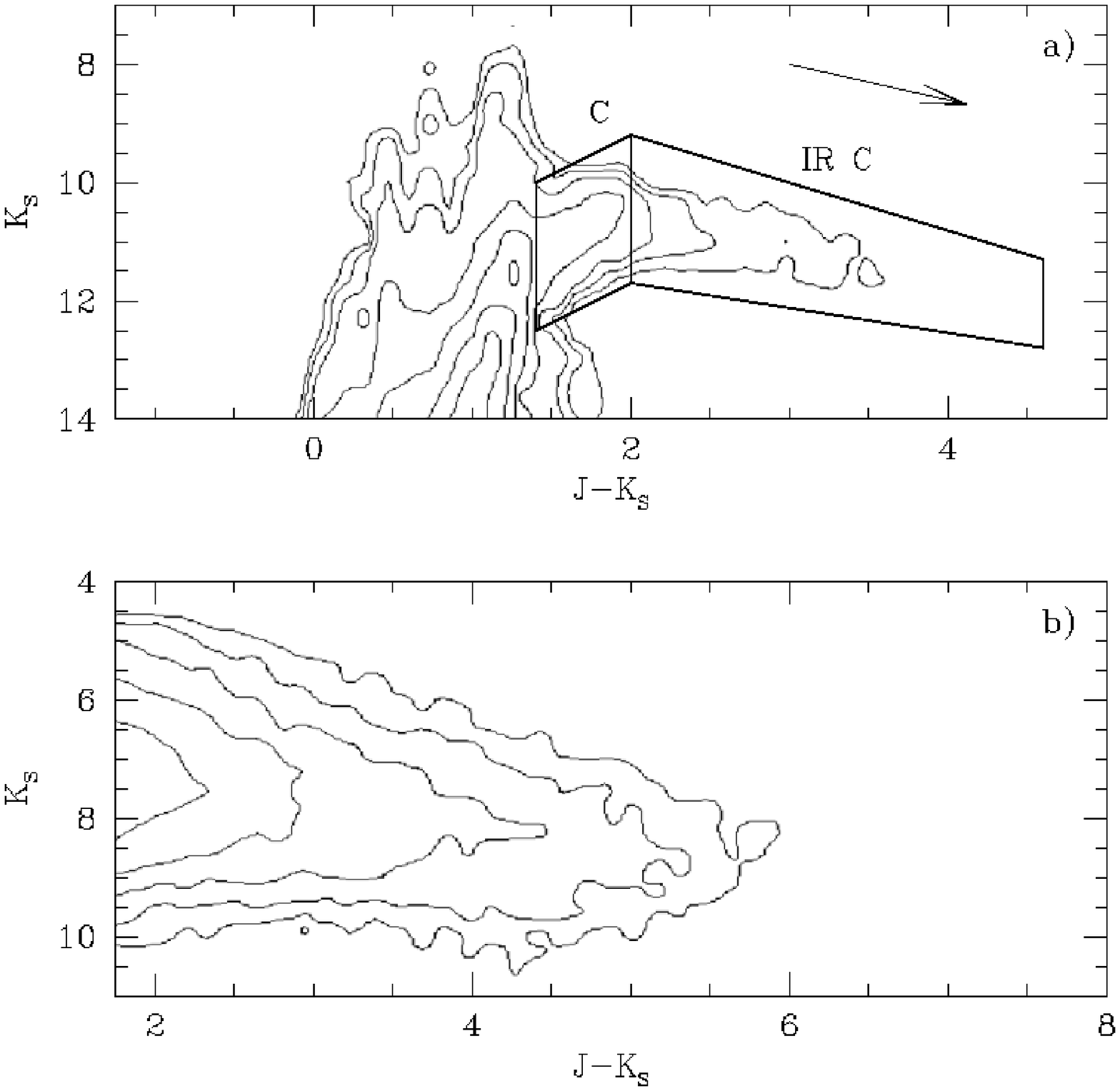}
\figcaption[figure1.eps]{2MASS Hess diagrams for: {\it a)} The
 LMC, showing the major stellar sequences and emphasizing the bilinear
 C~star sequence.  The arrow shows the reddening vector for
 E(B$-$V) = 2 mag, and {\it b)} The \irc\ sequence of the Milky Way,
 which shares a color-magnitude relation with that of the LMC and
 extends more than 2 magnitudes redder.  In the top (bottom) panel,
 each contour shows 4$\times$ (2$\times$) the density of the previous one.}
\end{figure}

\begin{figure}
\plotone{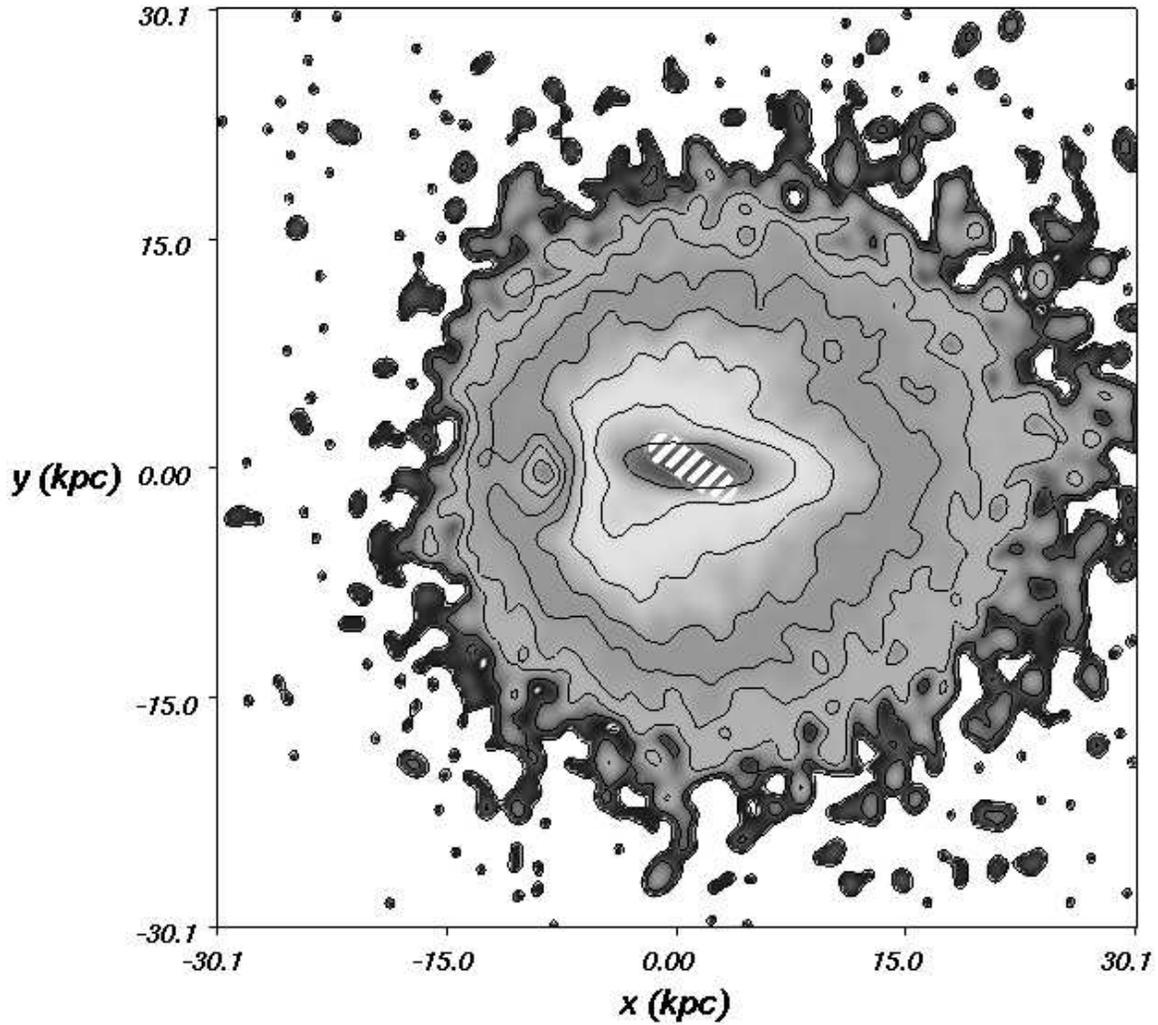}
\figcaption[figure2.eps]{Spatial distribution of \irc s within
 the Galaxy, with logarithmic contours.  The apparent hole 
 around the Sun ($x=-8$, $y=0$) owes to the saturation of
 nearby stars.  The symmetric disk structure and strong central
 bar are obvious; the position angle of the bar is distorted
 by the variance in \absk\ among the \irc s.  The hatched bar
 is our best estimate of the position angle and length,
 accounting for this bias.}
\end{figure} 

\begin{figure}
\plotone{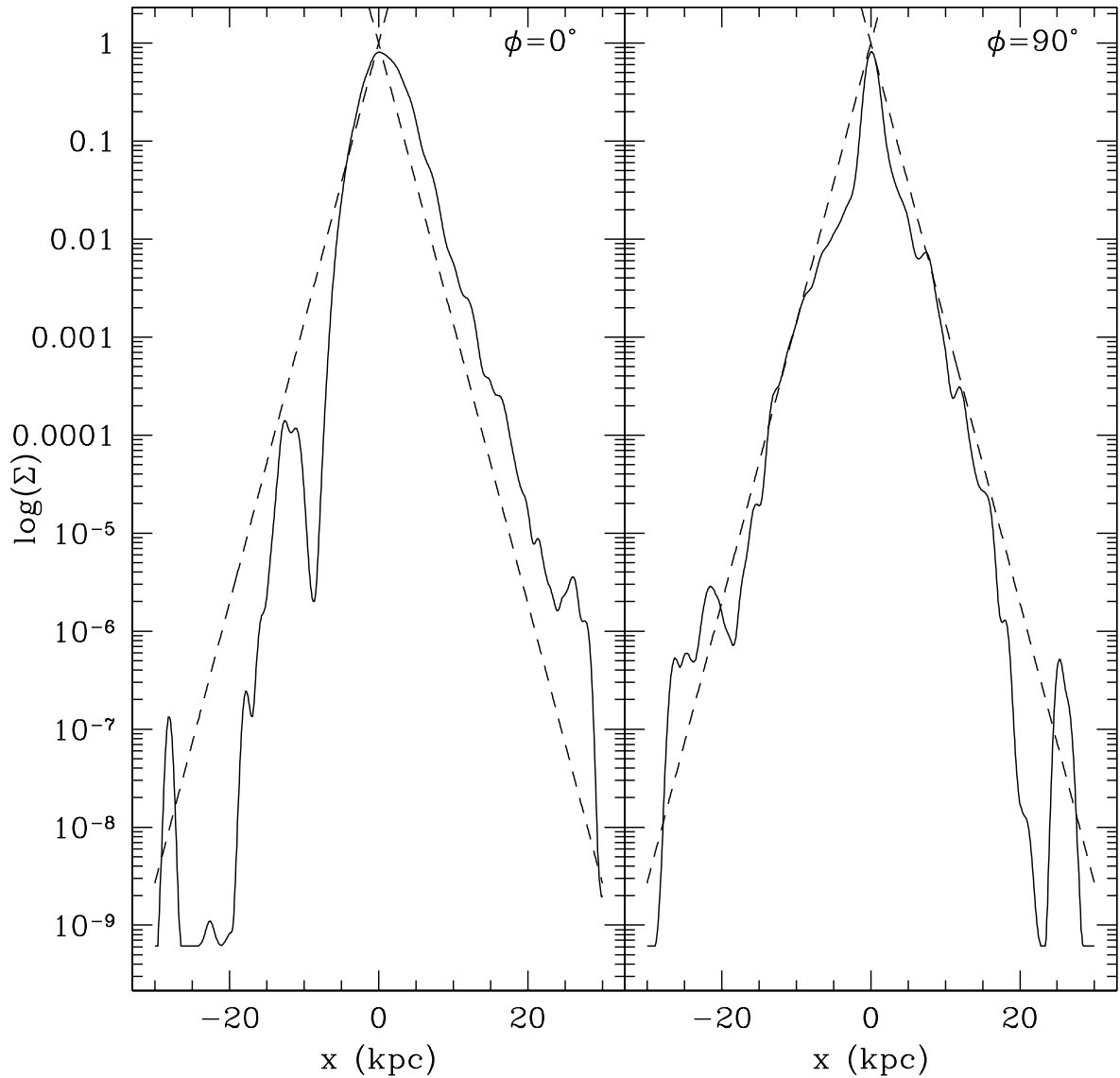}
\figcaption[figure3.eps]{Surface density profiles through the Galactic
  center parallel to the Sun-Galactic center line ($\phi=0^\circ$, 
  left) and perpendicular to it ($\phi=90^\circ$, right).  The 
  variance in luminosity for a given color (see Fig.~1) stretches
  the distribution for $\phi=0^\circ$.  The distortion is much smaller
  in the direction perpendicular to the line of sight        
  ($\phi=90^\circ$) and closely follows the inferred exponential
  profile of $a=3.5$ kpc out to the edge of the disk.  The dip at the
  Solar position, $x=-8$ kpc, is due to undersampling.}
\end{figure}

\begin{figure}
\plotone{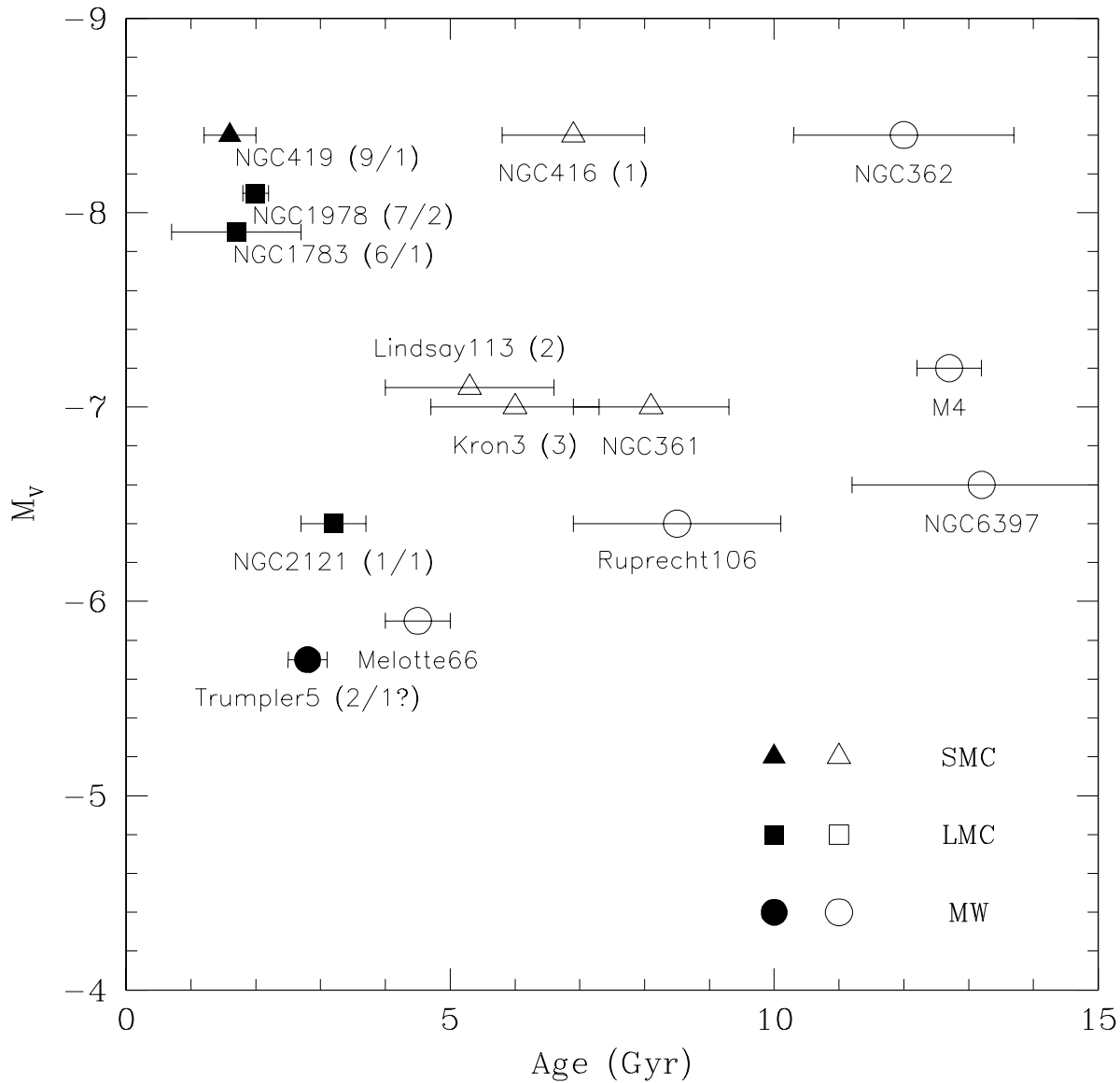}
\figcaption[figure4.eps]{Star clusters with and without C~stars.
  Clusters with known or suspected \irc\ members are marked as
  solid symbols, those without as open ones.  Globular cluster
  magnitudes are taken from \citet{har96}, with ages from 
  \citet{buo98}; MC and open cluster magnitudes are derived from
  SIMBAD data; their ages are from the recent literature, relying
  primarily on \citet{pia02}, \citet{nis00}, and \citet{sar99}.}
\end{figure}

\end{document}